\documentclass{pasj01}
\usepackage{mathrsfs} 
\usepackage{graphicx} 
\usepackage{mediabb}
\usepackage{color}  
\usepackage{bm}  
  
\title{Galactic Centre Threads as Nuclear MHD Waves}  
\author{Yoshiaki \textsc{Sofue}\altaffilmark{} }
\altaffiltext{}{Institute of Astronomy, The University of Tokyo, Mitaka, Tokyo 181-0015}


\begin{document} 
\date{ } 
\maketitle

\def\vlsr{v_{\rm LSR}} \def\Msun{M_\odot} \def\deg{^\circ} \def\/{\over}\def\kms{km s$^{-1}$} \def\Tb{T_{\rm B}} \def\sin{{\rm sin}\ } \def\cos{{\rm cos}\ } \def\Hcc{ H cm$^{-3}$ } \def\co{$^{12}$CO$(J=1-0)$ } \def\Htwo{H$_2$ } \def\be{\begin{equation}} \def\ee{\end{equation}} \def\({\left(} \def\){\right)} \def\[{\left[} \def\]{\right]}\def\Ico{I_{\rm CO} } \def\sech{{\rm sech}} \def\Htwocc{${\rm H_2\ cm^{-3}}$} \def\pr{p_r}\def\pt{p_\theta}\def\pp{p_\phi}\def\d{\partial}\def\cot{{\rm cot}} \def\Alf{Alfv{\'e}n } \def\muG{$\mu$G } \def\nH{n_{\rm H}}
\def\Vu{V_{\rm unit}} \def\rhou{\rho_{\rm unit}} \def\Bu{B_{\rm unit}} \def\tu{t_{\rm unit}} \def\Va{V_{\rm A}} \def\cs{c_{\rm s}}
\def\sub{\subsection} 
\def\Yu{Yusef-Zadeh }  \def\B{${\bm B}$ }
\def\rev{}

\begin{abstract} 
Propagation of fast-mode magnetohydrodynamic (MHD) compression waves is traced in the Galactic Center with a poloidal magnetic cylinder. MHD waves ejected from the nucleus are reflected and guided along the magnetic field, exhibiting vertically stretched fronts. The radio threads and non-thermal filaments are explained as due to tangential views of the waves driven by sporadic activity in Sgr A$^*$, or by multiple supernovae. In the latter case, the threads could be extremely deformed relics of old SNRs exploded in the nucleus.
\end{abstract}  
 
\section{Introduction}

Non-thermal filaments (NTF) and radio continuum threads (hereafter, 'threads') in the Galactic Centre (GC) are unique for their straight and narrow morphology perpendicular to the galactic plane
(\Yu et al. 1984, 2004; {Morris \& Yusef-Zadeh}{1985}; Tsuboi et al. 1986; Anantharamaiha et al. 1991; Lang et al. 1999a, b; LaRosa et al. 2004).
The wide-field image of the GC at 1.3 GHz with the MeerKAT (Heywood et al. 2019) has revealed a large-scale poloidal magnetic structure composed of coherently aligned vertical threads, which penetrate through the central molecular zone (CMZ: Oka et al. 1998, 2012; Tsuboi et al. 2015).
On the other hand, high-latitude threads are well correlated with the side edges of the Galactic Centre Lobe (GCL: Sofue and Handa 1984; Sofue 1985), which is one of the typical emerging phenomena commonly observed in the nuclei of disc galaxies.

There have been several ideas to explain the unique properties of threads and their physical relation to the central disk and nuclear activity (Sofue and Fujimoto 1987; Dahlburg et al. 2002;
 Boldyref and \Yu 2006; {Yusef-Zadeh \& Wardle} {2019}; {Barkov \& Lyutikov} {2019}). However, the mechanism to explain the straight vertical structures coherently penetrating the disk appears still unresolved. In this paper we propose a new mechanism for origin of the threads based on a simulation of the propagation of magnetohydrodynamic (MHD) waves excited by the activity of Sgr A$^*$ and supernovae around the nucleus. 
         
\section{MHD Waves and Thread Formation} 
 
Disturbances excited by explosive event in the nucleus propagate as a spherical shock wave in the initial phase. In fully expanded phase, they propagate as sound, \Alf, and fast-mode MHD waves of small amplitudes.  
The \Alf wave transports energy along the field lines, while it does not compress the local field. The fast-mode MHD wave (hereafter, MHD wave) propagates across the magnetic field lines at \Alf velocity, and compresses the local field, leading to enhanced synchrotron emission. 
The propagation of MHD waves originating at the nucleus is traced by solving the Eikonal equations developed by Uchida (1970, 1974), which has been also applied to the GC (Sofue 1977, 1980).  The equations are solved for a given distribution of \Alf velocity $V=\sqrt{B^2/4\pi \rho}$. 

We assume that the non-dimensionalized gas density is expressed by superposition of three components: (i) a plane-parallel disk with density distribution as
$\rho_{\rm disk}= \sech (z/h)$, where $h=1$ is the scale height, 
(ii) a molecular ring representing CMZ as
$\rho_{\rm ring}= 10 \ e^{-((r-r_{\rm ring})^2+z^2)/w_{\rm ring}^2}$,
where $r_{\rm ring}=5$ and $w_{\rm ring}=0.5$ are the radius and width of the ring, and 
(iii) a low density halo with $\rho_{\rm halo}=0.1$ 

The distribution of magnetic strength is expressed by an off-set, oblique, and open cylindrical form:
$B = [1.+\beta e^{-((r(x,y,z)-r_{\rm m})/w_{\rm m})^2}] [1.+(z/z_{\rm m})^2]^{-1}, $
where $ r(x,y,z)=[(x-0.5-0.2z)^2+(y-0.1z)^2]^{1/2}$, $r_{\rm m}=3(1+2(z/z_{\rm m})^2/(1+(z/z_{\rm m})^2))$, and 
 $\beta$ is a parameter to represent the magnetic strength, which is taken to be $\sim 10$, 
 $w_{\rm m}=1$, and $z_{\rm m}=3$. Here, $(x,y,z)$ are the Cartesian coordinates with $z$ being the polar axis, $r_{\rm m}$, $w_{\rm m}$ and $z_{\rm m}$ are the radius, width and vertical scale of variation of radius of the cylinder, respectively.
 
The real quantities are obtained  using the units of length $A$, time $\tu$, and velocity  $\Vu =\Bu/\sqrt{4\pi \rhou}$, and the following values are assumed: 
$\rhou=100$ \Hcc, $\Bu=1 $ mG, $A=20$ pc, leading to 
$\Vu=\Bu/\sqrt{4 \pi \rhou}=219$ \kms and $\tu=0.0898$ My ($\sim 10^5$ y).  
The \Alf velocity is calculated to be $\Va \sim 20$ \kms in the molecular ring and $\sim 200$ \kms in the galactic disc, which are higher than the sound velocities of the molecular ($\cs \sim 0.1-0.2$ \kms) and HI ($\sim 1$ \kms) gases. In the halo, $\Va$ increases rapidly due to the decreasing gas density to $\sim 10^3$ \kms or even higher in the magnetic cylinder, and is also higher than $\cs \sim 100$ \kms of the hot halo gas. In the present model, the entire region is thus assumed to be magnetically dominated, so that the Eikonal method can be safely applied.         

\begin{figure} 
\begin{center}   
\includegraphics[width=7.5cm]{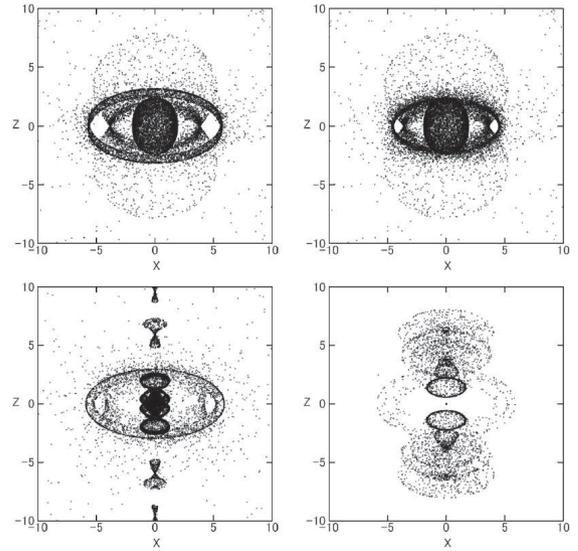}   
\end{center}
\caption{MHD wave fronts every $\delta t=2$ seen from $30\deg$ above the galactic plane through:  (top left) Sech disk with $h=1$: Waves make bipolar shells and converge onto a focal ring of radius 4.4. 
(top right) Same as a + gas ring of radius 5: Waves converge to the ring more efficiently. 
(bottom left) Same as b + magnetic cylinder of radius 3: Waves are guided by the cylinder.
(botom right) Same as c, but open magnetic cylinder: Waves are guided by the cylinder, but are more diverged and make bipolar caps.  }
\label{disk} 
\label{timeEvo} 
	\end{figure}

Figure \ref{disk}(top left) shows the result for a disk and halo with constant $B$. According to the rapid increase of $V$ toward the halo, the waves are strongly reflected and focus on a ring of radius $f_{\rm ring}\sim 4.4 h$. Here, the disc plays a role of a convex lens with a focal length of $f_{\rm disk}=2.2 h$ for a plane wave (Sofue 1977).
The top-right panel is a case when a gas ring of radius $r_{\rm ring}=5$ is added to the disk, where the \Alf velocity is lower than in the disk. The waves focus on and trapped by the gas ring.     
Figure \ref{disk}(bottom left) shows a case, when a straight vertical magnetic cylinder is added. The waves are strongly reflected by the inner wall of cylinder and are channeled through along the polar axis. Some fraction of the waves penetrate the cylinder and are trapped by the gas ring. The bottom-right panel shows the same, but the magnetic cylinder is open toward the halo. The waves are accordingly more open and make bipolar caps.  
  
\begin{figure} 
\begin{center}    
\includegraphics[width=6cm]{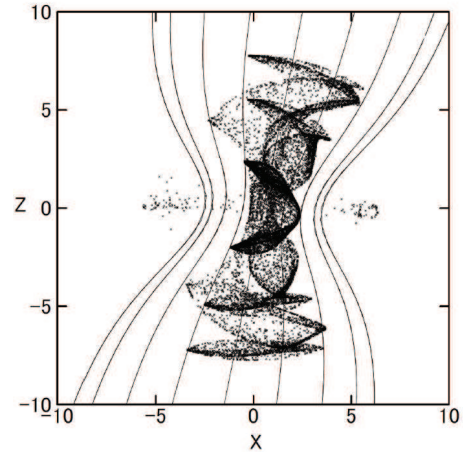}     
\end{center}
\caption{ MHD wave fronts seen from the galactic plane in an off-set, open, and oblique magnetic cylinder as drawn by the lines schematically. Waves are bent, deformed and stretched in the $z$ direction, while the both top fronts make bipolar caps.   } 
\label{timeEvoOb} 
	\end{figure}

\begin{figure} 
\begin{center}   
\includegraphics[width=6cm]{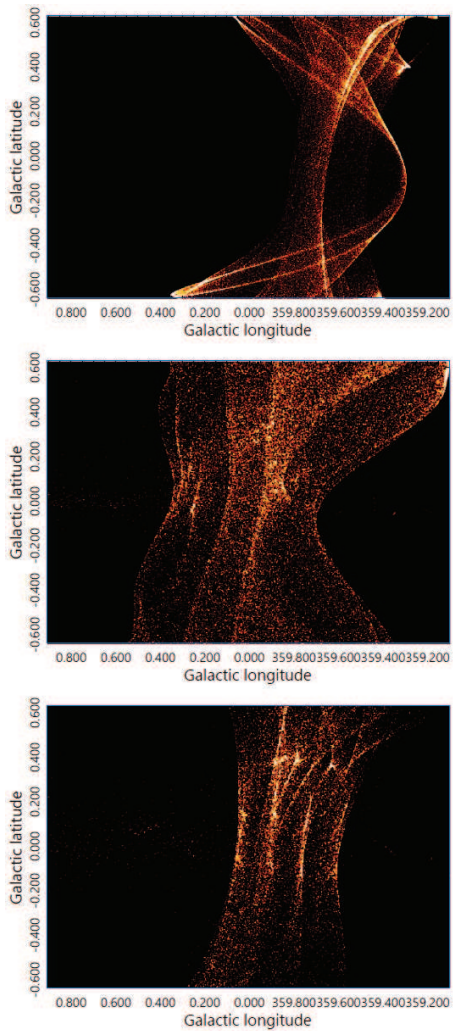}   
\end{center} 
\caption{MHD waves in an oblique, open and off-set magnetic cylinder at (top) $t=2-4$, (middle) $4-6$, and (bottom) $6-8$, exhibiting thread features as tangential projection of the wave fronts.
 Note that the figure does not represent the radio brightness. The radio emission is suppressed at high latitudes and in horizontal threads (see subsection \ref{radiobrightness} for radio emission).  }
\label{u-2-8} 
	\end{figure}

In order to simulate a more realistic case in the GC, an open, oblique, and off-set magnetic cylinder is added to the disk and ring, and the result is shown in figures \ref{timeEvoOb}. The left-side half is reflected by the near-side inner wall of the cylinder, and expands backward toward the other side. It, then, merges with the original half, and both are guided along the cylinder to result in a bipolar channeled MHD waves. The front makes finally a vertically stretched and corrugated sheet near the disk and bipolar caps in the halo. 
Figure \ref{u-2-8} enlarges the central region at $t=2-4$,  $4-6$, and  $6-8$.
  Note that the figure does not represent the radio brightness, which is much suppressed at high latitudes and in horizontal threads (subsection \ref{radiobrightness} for radio emission). 
   
\begin{figure} 
\begin{center}               
 \includegraphics[width=7cm]{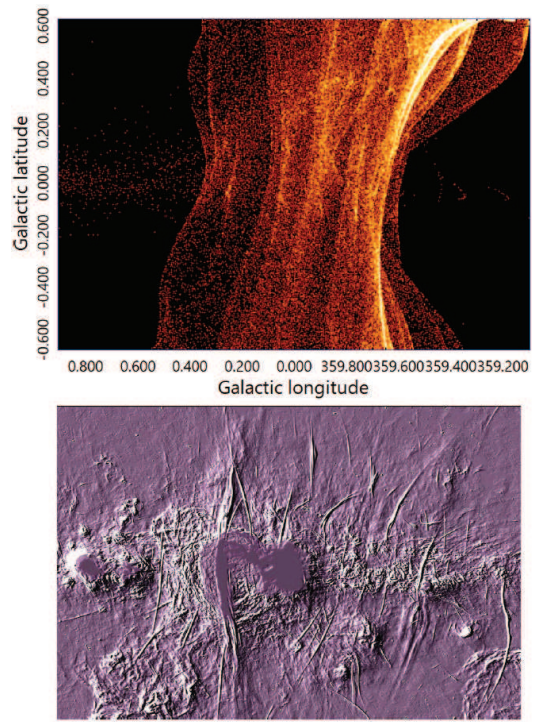}  
\end{center} 
\caption{Projection of 3D MHD wave fronts on the sky at $t=3$ to 8 by interval $\delta t=0.5$ (for radio emission, see subsection \ref{radiobrightness}). Bottom panel shows a 1.3 GHz radio image taken from the web page of NASA (https://apod.nasa.gov/apod/ap190708.html, credit: MeerKAT, SARAO) (Heywood et al. 2019). Threads are enhanced in relief.  }
\label{model-vs-mKAT} 
\label{mKAT} 
	\end{figure}

Figure \ref{model-vs-mKAT} shows the fronts from $t= 3$ to 8 at smaller interval 0.5, and compares with the observation at 1.3 GHz (Heywood et al. 2019). 
The tangential projection of the MHD front sheets well mimic the observed radio threads. Parallel magnetic orientation in isolated threads (Lang et al. 1999a)  may be understood as due to the projected view of compressed field lines in the fronts. Again, see subsection \ref{radiobrightness} for radio emission.

\section{Discussion}

\sub{Poloidal magnetic field}

\begin{figure} 
\begin{center}  
\includegraphics[width=8.5cm]{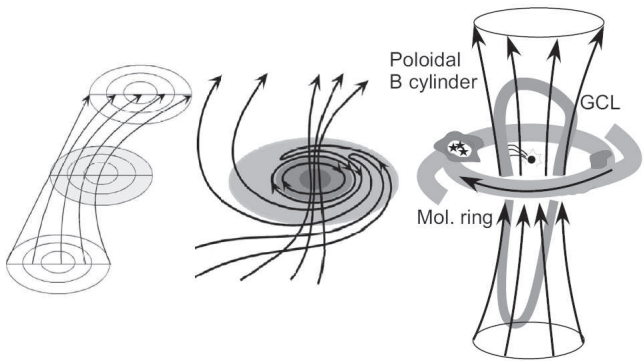}  
\end{center}
\caption{Formation of poloidal magnetic field in the GC, ring and spiral fields in the disk (Sofue and Fujimoto 1987; Sofue et al. 2010). Right panel illustrates the GCL, molecular ring (CMZ), and supposed poloidal field in the GC. }
\label{illust_primo} 
	\end{figure}

 It has been shown that nuclear poloidal field is created by secular accumulation of intergalactic magnetic field frozen into a forming galaxy (Sofue and Fujimoto 1987; Sofue et al. 2010). Since the vertical component can neither escape across the accreting disc, nor be dissipated because of the absence of neutral sheet, they are secularly concentrated onto the GC and form a strong vertical flux (figure \ref{illust_primo}; Sofue and Fujimoto 1987). The model is in accordance with the radio continuum observations of the well ordered poloidal magnetic structure penetrating through the GC disc without twisting (Heywood et al. 2019).

The vertical field is maximized in a cylinder around the polar axis, being sandwiched between the accumulating flux from outside and high pressure from inside by the hot nuclear gas (Nakashima et al. 2019; Ponti et al. 2019).
 Since the magnetic field is anchored to the intergalactic space at rest, the accumulated field will be also at rest, when it became stationary after a cosmic time. This means that the field lines run through the cavity inside the CMZ's ring without rotation. 

If the magnetic field is not rotating, the frozen-in gas would be also at rest,  which has been indeed confirmed by the recent observation of slow or almost no rotation of the GCL using the H92$\alpha$ recombination-line (Nagoshi et al. 2019). The coherently aligned, straight, and non-twisted morphology indicates that the threads are not interacting with the rotating gas disc. The threads are further continued by an even longer radio spur at 2.7 GHz reaching to $z\sim -350$ pc (Reich et al. 1984). All these facts suggest that the threads are a part of a structure at rest continued from the intergalactic space. 
 
\sub{Proper motion}

Although the magnetic field is at rest, threads are moving at the \Alf velocity of the order of $\sim 200$ \kms at random, so at mutual random velocities of $\sim 400$ \kms. This would cause proper motions between neighboring or overlapping threads on the order of $\sim 10 \ {\rm  mas \ y^{-1}}$. On the contrary, if the magnetic field is rotating with the disc, and the filaments are thin magnetic fluxes, the threads' proper motion would be more systematic in that the mutual proper motion decreases with the distance from the rotation axis.  

\sub{Magnetic strength and radio brightness}
\label{radiobrightness}

Let $\Delta$ be the thickness of an MHD wave front and $R$ the curvature. Then the tangential line-of-sight depth is given by $\Lambda_{\rm wave}=\sqrt{2R\Delta}$. If we assume $R\sim 30$ pc and $\Delta \sim 0.04$ pc ($1''$) as for the thinnest thread (Par{\'e} et al. 2019), we obtain $\Lambda_{\rm wave} \sim 1.5$ pc. On the other hand, if the thread is a string, the depth would be about the same as its diameter, $\Lambda_{\rm string}\sim \Delta \sim 0.04$ pc. Then, the ratio of emissivity required for the observed radio brightness by the MHD wave model to that for the string model can be reduced by a factor of  $\kappa =\epsilon_{wave}/\epsilon_{\rm string}=L_{\rm string}/L_{\rm wave} \sim 0.03$ .  
 
Assuming the energy equipartition for a volume emissivity $\epsilon\sim \nu \Sigma_\nu/\Lambda$, where $\Sigma_\nu \sim 80$ mJy/beam by $\sim 1''$ beam at $\nu=10$ GHz (Par{\'e} et al. 2019) and measured thread width of $\sim 1''$, the magnetic strength is estimated to be $B\sim 0.6$ mG in the string model, whereas $B\sim 0.2$ mG in the wave model. The wave model can, thus, save the energy by an order of magnitude compared to string model. 
 
The radio emissivity, and hence the brightness $\Sigma$, varies with $B$ as well as with the angle between the directions of wave propagation and field, $\theta$. We may then approximate the variation of synchrotron brightness as $\Sigma \propto (B_{\perp})^{3.5}\sim (B \ \sin \theta)^{3.5}\propto [(1+(z/z_{\rm m})^2)\sin \ \theta]^{-3.5} $. A model radio map may be obtained by multiplying this factor to the front's projected density maps. This results in a more rapidly decreasing radio brightness with latitude and suppressed emission for horizontal features. The larger number of brighter threads near the galactic plane may be understood by such latitudinal as well as the $\theta$ variation of the emissivity.

 \sub{Dissipation and heating of GCL}
 
Neglecting the Ohmic loss, the dissipation rate $\gamma$ of the MHD wave (Landau and Lifshits 1960) is given by
$
\gamma=\omega^2 \nu_{\rm d}/(2V^3\rho),
$
where $\omega=2 \pi V/\lambda$ is the frequency and  $\nu_{\rm d} $ is the viscosity of hydrogen gas. 
The dissipation length, $D=1/\gamma$, is then estimated by
$(D/pc)\sim 570 (B/\mu{\rm G})$
$(\rho/{\rm H \ cm^{-3}})^{1/2}$
$(\nu_{\rm d} /10^{-4}{\rm g\ cm\ s^{-1}} )^{-1}$
$(\lambda /{\rm pc} )^2$. 
This yields sufficiently long distance, $D\sim 5$ kpc, near the galactic disk for $B\sim 1$ mG and $\rho\sim 100$ \Hcc, and the wave is dissipation less. 

On the other hand, outside the disk with less density and magnetic field as $B\sim 0.1$ mG and $\rho\sim 1$ \Hcc, wave is dissipated in $D\sim 50$ pc, and thermalized to heat the gas inside the magnetic cylinder. According to the decrease in the gas density and magnetic strength, the wave amplitude will increase with the height from the Galactic plane, and at some height, they will be rapidly dissipated to heat the surrounding gas.

It may be also mentioned that a comparable fraction of the energy released in the nucleus is transported in the form of \Alf waves in the polar direction along the vertical magnetic fields. According to rapid decrease in the halo gas density, the wave amplitude will increase, and attain non-linear growth at a certain height, where the waves will be dissipated to heat the GCL.

Such heating of the halo gas due to MHD- and \Alf-wave dissipation as well as accumulation of the wave flux at high latitudes would become a heating source of the GCL, which is observed to be filled with ionized hydrogen gas (Sofue 1985; Nagoshi et al. 2019). 

\sub{Unique morphology}

\begin{figure} 
\begin{center}        
 \includegraphics[width=6cm]{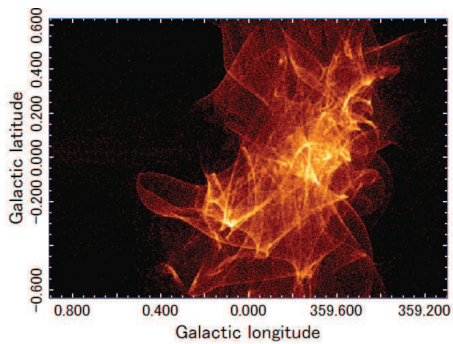}  
\end{center} 
\caption{MHD wave fronts in an off-set and oblique straight magnetic cylinder through a gas disk with tight concentration to the plane, resulting in strong confinement near the plane and complicated structures.  
 Note, however, that the synchrotron radio brightness of the wave front decreases with the latitude more rapidly and horizontal features are much suppressed (see subsection \ref{radiobrightness}).}  
\label{u-extreme} 
	\end{figure}

\begin{figure} 
\begin{center}          
 \includegraphics[width=6cm]{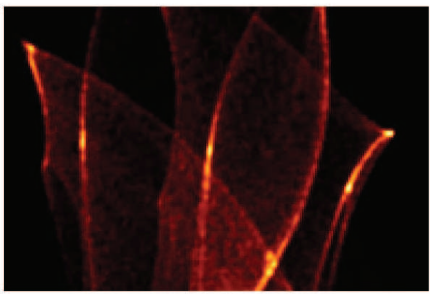}   
\end{center} 
\caption{Enlargement of a portion of MHD waves, showing bifurcation, kink, bend, mice, and crossed threads.  }
\label{kink} 
	\end{figure}

The present model well reproduces not only the straight filaments, but also those with peculiar morphology in the Arc (Par{\'e} et al. 2019) and horizontal threads (Lang et al. 1999b):
Bifurcated and crossed threads are due to projection of two or multiple fronts ejected at different epochs and/or those superposed by retarded reflections. Double and multiple strings are explained also by such multiple fronts. Threads having kinks, bends, and bright knots, and maybe mouse-like spots, can be explained by superposition of obliquely corrugated and warped MHD fronts. 
Figure \ref{u-extreme} shows an example of simulation in a straight and off-set magnetic cylinder and a disk with smaller scale height, where the increase of \Alf velocity is steeper, so that the reflection is stronger. The waves are confined near the origin, and exhibit a variety of filamentary structures. Figure \ref{kink} enlarges a part of the simulation, showing crossed and horizontal threads as well.

\begin{figure} 
\begin{center}         
 \includegraphics[width=6cm]{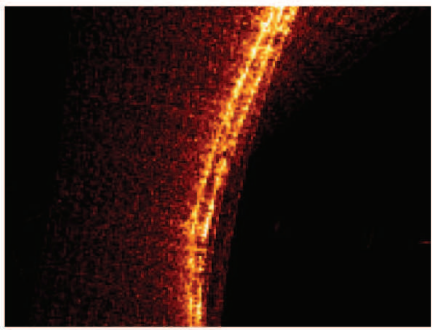}      
\end{center} 
\caption{Enlargement of a portion of MHD waves at $t=3$ to 4 by interval of $\delta t=0.1$, mimicking the NTF in the Arc. Faint horizontal waves are superposed, making the threads apparently clumpy.  }
\label{t3} 
	\end{figure}

The bunched NTF in the radio Arc (\Yu et al. 1984) could be due to a past temporal enhancement of activity associated with multiple disturbances at shorter time interval. Figure \ref{t3} shows an enlarged portion of the simulated waves from $t=3$ to 4 at an interval of $\delta t=0.1$, where the tangential projection of the fronts resembles the observed NTF in the Arc. It is stressed that the threads show apparently clumpy features due to superposition of horizontal and fainter waves in the fore- and backgrounds, which is indeed observed (Par{\'e} et al. 2019). 

\sub{Time variation in Sgr A}

Sporadic arrival of the waves produces multiple threads, whose spatial interval manifests time interval of the ejections as $\delta L\sim V \delta t$.   
The interval between the brightest threads is observed to be $\delta L\sim 20$ pc (e.g., Anantharamaiah et al. 1991), which corresponds to time interval of $\delta t \sim 10^5$ y for $V\sim 200$ \kms. This time may be compared with those suggested for the GCL ($\sim 0.1$ My), Fermi Bubbles (FB: $\sim 1$ My), and the bipolar hyper shells (BHS: $\sim 10$ My) (Kataoka et al. 2017 for review). 

Numerous minor threads at interval of $\delta L\sim 2$ pc (Heywood et al. 2019; Per{\'e} et al. 2019) may correspond to waves emitted in shorter time scales of  $\sim 10^4$ y. Furthermore, much shorter interval and fainter threads may exist according to less energetic and shorter-time activity as observed as the time variation of Sgr A$^*$ ({Subroweit et al.} {2017}). Such weak waves could be observed as numerous faint threads with sub- to milli-pc intervals.
 
\sub{{\rev Possible origin of the threads and} energetics} 

The magnetic energy contained in a single {\rev thread} is estimated to be $E\sim B^2/(8 \pi) 4 \pi R^2 \Delta \sim 10^{50}$ erg for assumed magnetic strength of $B\sim 0.3$ mG (log mean of 0.1 to 1 mG) with $R\sim 30$ pc and $\Delta \sim 0.04$ pc. Namely, about 10\% of released kinetic energy by a single core-collapse supernova (SN)  ($10^{51}$ erg) is sufficient to drive one thread. This energetic proximity might indicate that a thread is an extremely deformed relic of an old SNR. Namely, the presently observed threads might be an ensemble of such old SNRs due to a bursting SNe in the nucleus some $10^5$ y ago. 
Alternatively, the energy source could be sporadic puffing activity of Sgr A$^*$. This idea applies particularly to numerous fainter threads, which may be waves driven by less energetic activities in the nucleus with shorter time scales. 
In either case, the total energy to drive all the observed threads is on the order of $\sim 10^{52}$ erg. Or, only a small fraction of energy required for GC bubbles like GCL, FB, and BHS ($10^{54}-10^{55}$ erg) will be sufficient to produce the GC threads.

\section{Summary}

Propagation of fast-mode MHD compression waves was traced in the GC penetrated by an open, off-axis and oblique poloidal magnetic field. 
Reflected and guided waves exhibit vertically stretched fronts, and their tangential projections well mimic the morphological properties of radio threads and non-thermal filaments.
The origin of the poloidal field is explained as due to secular accumulation of frozen-in primordial magnetic fields to the Galaxy.
Energetics suggests that the MHD waves are due either to sporadic puffing in Sgr A$^*$ or to multiple SNe some $10^5$ y ago. 
In the latter case, the threads may be regarded as an ensamble of extremely deformed relics of old supernova remnants exploded in the nucleus. 

 \vskip 1mm 
\noindent{\bf Aknowledgements:}  
Computations were carried out at the Astronomy Data Center of the National Astronomical Observatory of Japan.

\end{document}